\documentclass{article}
\usepackage{emsnewsletter}
\newtheorem*{law*}{\textbf{Copernicus-Gresham Law}}
\newtheorem*{corollary*}{\textbf{Corollary}}
\newtheorem*{theorem*}{\textbf{Transparency Theorem}}

\author{Ricardo P\'erez-Marco (CNRS, Univ. Paris 13, Paris, France)}
\title{Bitcoin and Decentralized Trust Protocols}

\begin{document}
   
\maketitle

\begin{abstract}
Bitcoin is the first decentralized peer-to-peer (P2P) electronic currency created in November 2008 by 
Satoshi Nakamoto, the pseudonym of the anonymous author or group of authors of the groundbreaking article \cite{Nakamoto08}. 
Moreover, Nakamoto released the first implementation of the protocol 
in an open source client
software \cite{Nakamoto09} and the genesis of bitcoins began on January 9th 2009. 
The Bitcoin protocol is based on clever ideas which solve a form of the Byzantine Generals Problem and sets 
the foundation for Decentralized Trust Protocols. Still in its infancy, the currency and the protocol have 
the potential to disrupt the international financial 
system and other sectors where business is based on trusted third parties. 
The security of the bitcoin protocol relies on strong cryptography and one way hashing algorithms.  
\end{abstract}

\section{Electronic decentralized money}\label{sec:intro}  

Progress in cryptography has made possible secure communications and electronic payments over the Internet. 
The use of a credit card is a form of electronic cash which relies on a trusted third party, preventing 
overspending or double spending. The remarkable main contribution of the Bitcoin protocol is to
get rid of trusted third parties and establish a pure peer-to-peer (P2P) decentralized currency.

\medskip

Our national currencies rely on the central 
banks that have the power to regulate the monetary mass. 
These central authorities are not always public institutions, like the Federal Reserve. 
In the last decade the European Central Bank (ECB) 
has taken over national central banks in the European Union.
The ECB claims its independence from national governments in order to set monetary policy. 
Unfortunately the role of the ECB is not politically neutral as seen in 
recent Greek crisis. The elected Greek government come under pressure from the ECB 
when deprived Greek banks of liquidity. Whoever owns the key to the printing press has 
tremendous economic and political power.

\medskip

A decentralized form of money has existed for several thousand years. Precious metals, gold and silver, are decentralized: 
Value is widely acknowledge because of scarcity and physical properties. The ``barbarous relic''(quoted 
from J.M. Keynes making reference to the gold standard) 
has been the ``canonical'' form of money since ancient times. From cuneiform clay tablets \cite{CDLI} we know that precious metals
metals were used in Mesopotamia already circa $2600$ B.C.
There was an evolution in Europe during the $18^{\rm th}$ century when bank notes were introduced by John Law 
(earlier paper or leather money appeared in China \cite{ServalTraine2014}). 
These were backed by gold and silver (and land).
Before modern times we had world monetary bimetalism (during the Roman Empire Aureus 
and Denarius coins were used). This generates monetary tensions due to the fluctuations of one metal 
with respect to the other, but it gives a subtle monetary equilibrium (see \cite{Flandreau2004}) that 
can be disrupted by fixing the wrong exchange rate. This happened in 1717 when Isaac Newton, then Master of the British Royal Mint, 
set an incorrect gold/silver ratio imposing a de facto gold standard.
After 1870 with the withdrawn of the stabilizing role of the Banque de France (unable to sustain the gold/silver ratio at $15.5$. A 
$20$ French Franc gold Napoleon coin contains $5.81$ grams of pure gold and the $5$ French Franc silver coin $22.5$ grams of 
pure silver), we assist to the emergence of a world gold standard: Gold becomes the world currency. 
National currencies are then backed by gold reserves.  The dynamics between competing currencies obeys the \textbf{Copernicus-Gresham Law} (stated 
by Copernicus \cite{Copernicus1526}, and earlier by Oresme \cite{Oresme1360}): 
\begin{law*}
\textit{Bad money drives out the good.} 
\end{law*}
This means that if two kinds of money are available, people prefer to use or spend ``bad'' money, and hoard ``good'' money. 

Historically the gold standard was gradually abandoned during the $20^{\rm th}$ century, until 1971 when the convertibility of the American dollar 
into gold was withdrawn (Nixon shock) in favor of a floating currency system. Then national currencies became ``fiat'' money 
backed only by the faith in the issuing central banks and governments.
  
%

\begin{figure}
\centering
\includegraphics[width=.8\linewidth]{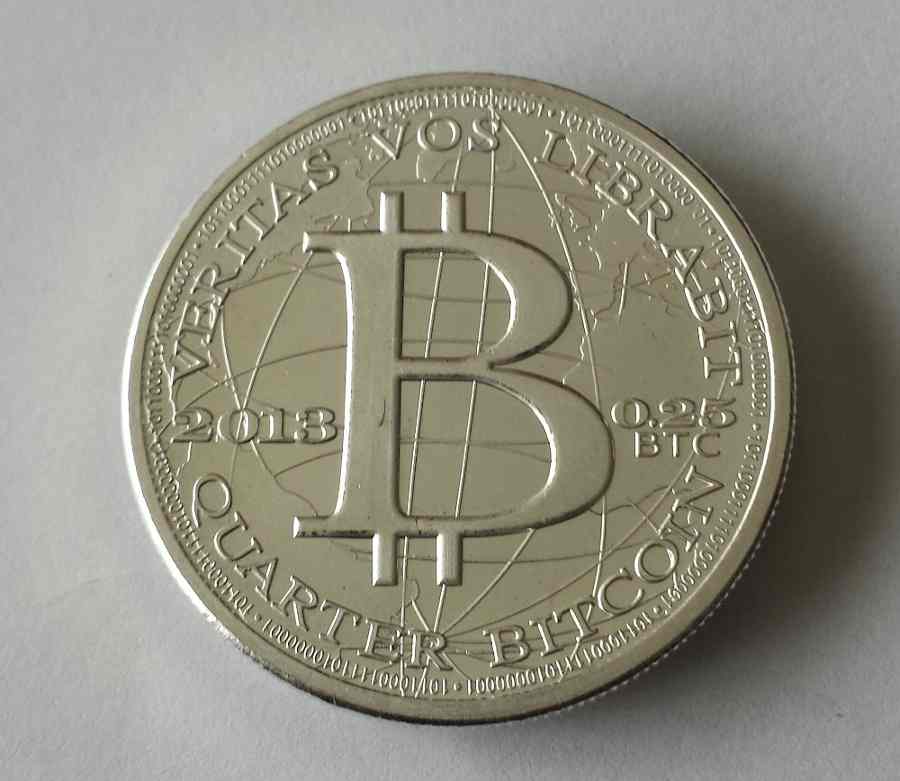}

\medskip

\caption{Old and new decentralized money blended together into a one ounce silver coin.}\label{fig:bitcoin}
	
\end{figure}

The motivation behind Bitcoin creators is to create a form of ``electronic gold'' whose integrity and non-falsifiability relies on mathematical 
properties, instead of physical properties for gold, or faith on central banks for fiat money. 
How is this even possible when a digital token can be replicated exactly, infinitely, at no cost? It would be like being able to produce gold
easily, jeopardizing the scarcity and non-falsifiability properties that make it valuable. 
But on the other hand, the electronic 
nature makes it perfect for storage and transportation. The main obstacle we face is to prevent the possibility of "`double spending"', 
i.e. the simultaneous use of the same token for different payments. 
At first, ``decentralization'' and ``electronic'' seem to be incompatible goals. The 
``double spending problem'' is the main difficulty to the inception of decentralized electronic money. 

The only way to prevent double spending is to have a ledger accounting for all transactions, so that the recipient can check that the 
transaction is legitimate. If we don't want this ledger to be centralized 
under the control of a third party, then it must be public. This argument shows that all transactions must be publicly recorded (maybe in an obfuscated form).
 
\begin{theorem*}
\textit{Electronic decentralized money must rely on a public ledger.}
\end{theorem*}

Once this is acquired, the major problem consists in how to construct a trusted public ledger, 
that is, how to build a reliable non-falsifiable public database of approved transactions. This is not a simple task. 
For bitcoin this public ledger is a file or a set of files called ``the blockchain'', which contains indeed a chronological 
sequence of blocks cryptographically bundled together with all bitcoin transactions.

\section{Byzantine Generals and Sybil attacks.}\label{sec:byzantine_generals}

At the core we need to establish a decentralized mechanism for trust consensus. A centralized ``trust'' stamp is as good as the 
confidence put in the third party which validates trust. Therefore decentralized trust consensus, if possible, is stronger 
and more resilient. This is a good example of ``Taleb antifragile structure'' \cite{Taleb2012} since it does benefit from the erosion of centralized 
systems. It is an intricate (byzantine!) problem to devise such a mechanism of validation. 
This ``Trust Machine'' (as presented in the cover page of a recent issue of ``The Economist'' \cite{TheEconomist2015}) is the core of the Bitcoin protocol. 

We cannot discard that some agents may act maliciously, and we can neither assume that the communications are secure, 
nor control who participates in the open community.

This type of problem was first systematically studied since 1982 \cite{LAMPORT:1982}. Lamport named it 
``The Byzantine Generals Problem'' in the context of computer systems. The goal is to handle malfunctioning or malicious components which 
give conflicting information to different parts of the system. As these authors present it:

\medskip

\textbf{The Byzantine Generals Problem.}

\textit{
The situation can be described as the siege of a city by a group of generals of the Byzantine 
army. Communicating only by messenger, the generals must agree upon a common battle plan. However, one or more of them may be traitors 
who will try to confuse the others. The problem is to find an algorithm to ensure that the loyal generals will reach an agreement. 
}

\medskip

The problem of reaching consensus in an open network is similar, but more complex: The number of generals is not fixed. 
In this more general ``Nakamoto Byzantine Generals Problem'' we must 
prevent a pseudospoofing or Sybil attack which consist in the creation of multiple fake identities in order 
to subvert the reputation of the system. The safeguard against this form of attack is to request that the participation or share of influence 
in the system has a cost. Thus a malicious attacker would need important resources. 
Nakamoto's proposal is to make the influence of each participant proportional to the share of computer 
power that he contributes to secure the system. The idea of ``Proof of Work'' (PoW) was used by Adam Back with \textit{Hashcash} (1997)
to fight spam. The first implemented cryptocurrency using a PoW was Hal Finney RPOW. Other earlier proposals were Wei Dai with \textit {b-money} (1998) and 
Nick Szabo with \textit{Bit Gold} (1998). According to Nakamoto (bitcointalk 2010) \textit{``Bitcoin is an implementation of Wei Dai's b-money proposal on Cypherpunks in 1998 and Nick Szabo's Bitgold proposal''}.

The network is secure as long as the majority of the computer power comes from honest players. 

\section{Nakamoto decentralized consensus.}

\begin{figure}
\centering
\includegraphics[width=1.08\linewidth]{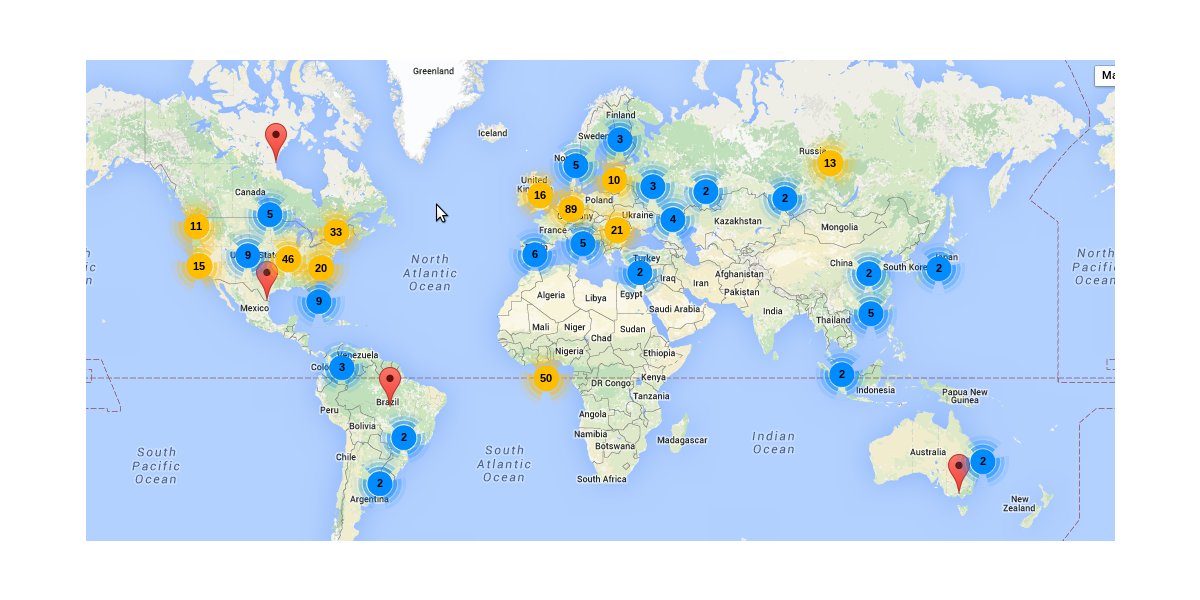}

\caption{Worldwide propagation of a bitcoin transaction.}\label{fig:propagation}
	
\end{figure}

The process to validate bitcoins transactions safeguards against double spending works as follows. The Bitcoin net consists in nodes that connect P2P to each other 
and propagate the transactions (as in figure \ref{fig:propagation}). The typical client connects to at least $8$ other peers. 
Each node checks that the transactions are valid and respect the rules of the protocol (the code is public 
\cite{github}). Some of these nodes are ``miners'' or validators of transactions.  
They collect transactions into a 
block of transactions. A hashing algorithm diggest any file into a fixed length string of bits 
(more on hashing algorithms below). 
Miners add the hash of the header of the previously 
validated block and some other data (timestamp, number of transactions,...) in order to build the new block header. 
Varying a nonce they try to find a hash of the block header  
starting with a certain number of zero bits. This number sets the difficulty 
of the problem and is adjusted every $2016$ blocks in function of the total hashing power so that a solution is found in about $10$ minutes on average. 
This procedure is a sort of ``decentralized lottery'' that designates the miner that validates the blocks. The ``winner'' earns a reward in newly created 
bitcoins. The winning probability is proportional to the coputer power provided. The validated block is propagated through the network that checks 
that the solution is correct and the block does not contain double spends.

The miners hash the header of the block but not the whole block. To insure incorruptibility of the transactions, the root of a Merkle hash tree
(more on Merkle trees below) 
of the transactions is added to the header.

A race starts when two or more blocks are validated simultaneously in different parts of the net. Then some miners accept one and others the other 
one as the next valid block. In further validations one of the two groups will be faster. Then all miners switch to the longest 
blockchain and the race stops. This creates what are called ``orphan blocks'' that are off the blockchain and invalid. 

The miner validating the block gets a reward in bitcoins. This was first set to be $50$ bitcoins and the reward is halving in about every $4$ years (more precisely,
after $210.000$ blocks). At the time of writing the reward is of $25$ BTC, and by the end of July 2016 after block number $420.000$ the reward 
will drop to  $12.5$ BTC.

Apart from this mining reward, miners can get fees from each transaction. Fees are voluntary, up to the user, 
and typicaly zero or very low. But a higher fee incentives the miners to 
include the transaction with priority in the next block. Thus a higher fee ensures that the transaction are processed faster.  

The production of bitcoins slows down in a geometric way. At the time of writing the total number of bitcoins in existence is over $15$ millions. 
By year 2140 a total of $21$ million bitcoins will exist and no more will be produced. The growth of the monetary mass is programmed into the algorithm. Each bitcoin 
can be fractioned in $100$ million units called ``satoshis'', i.e. we can use 8 digits. This makes the currency extremely divisible and suited for micropayments.
The model of production and the scarcity imposed is inspired from mining gold.

\subsection*{Hashing algorithms.}

A hashing algorithm digest any file into a fixed length string of bits. The slightiest modification of the original file produces a completely different output.
The bits of the output appear with a random frequency and it is computationnally hard to find collisions (different inputs yielding the same output).
Hashing algorithms are used for example to check the integrity and non-tampering of files. 

The two main hashing algorithms used in the Bitcoin protocol are RIPEMD-160 and SHA-256 that produce outputs of $160$ and $256$ bits respectively. The mining algorithm 
consists in performing the double SHA-256 of the block header (doubled to prevent ``padding attacks''). 
The Merkle tree of transactions in a block consists in pairing the transations and hashing them together 
(repeating one in order to get a power of $2$), then pair the resulting hashes, etc in order to build a dyadic 
tree of hashes. The root of the tree is the Merkle root and is included into the header of the block. 
So we cannot change a transaction nor their ordering once the block is validated.

\section{Pseudoanonymity and structure of bitcoin addresses. }

Since transactions must be made public, it seems that all privacy is lost.
This is not the case because bitcoin addresses can be created without any personal data from the owner.

Each bitcoin address is composed by a pair of public and private keys. The private key is used to sign transactions that  
allow to spend the bitcoins in the balance of the address. Technically speaking the owner of the bitcoins in that address 
is whoever has control of the associated private key. The public key has a shorter hashed version that is the one made public in order
to receive funds. A bitcoin public address looks like \\
$$
\scalebox{0.95} {14xuSZXtfGw5XqfYxEjp4crwYGYQDWmZ12}
$$
This one is under the control of the author (and you are welcome to try your hand at bitcoin transactions by sending many bitcoins!). 
With a QR code encoding your address you can scan it with a smartphone.

\begin{figure}
\centering
\includegraphics[width=.6\linewidth]{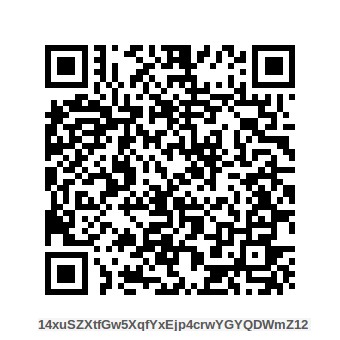}

\caption{QR code of a bitcoin address}\label{fig:QRcode}
	
\end{figure}

Bitcoin addresses are generated using Elliptic Curve Cryptography. To generate a private key, 
we pick randomly $256$ bits (by tossing a coin $256$ times for example),  and this binary number 
gives a secret key $k$. 
The associated public key is computed using elliptic curve multiplication $k.g$ of a base point $g$ in the \textit{secp256k1} 
elliptic curve
$$
y^2=x^3+7
$$
where the arithmetic is done in the prime field $\mathbb{F}_p$ where the prime number $p$ is slightly less than $2^{256}$,
$$
\scalebox{0.9}{$p=2^{256}-2^{32}-2^9-2^8-2^7-2^6-2^4-1$} \  ,
$$
and the base point is given by
\begin{align*}
g_x&\scalebox{0.6}{=55066263022277343669578718895168534326250603453777594175500187360389116729240}  \\
g_y&\scalebox{0.6}{=32670510020758816978083085130507043184471273380659243275938904335757337482424} 
\end{align*}
This choice of elliptic curve and base point follows cryptographic standards (which is the subject of some controversy for the unexplained 
choice of parameters). The public key is the $520$ bit number obtained as $K=04xy$ by concatenating the hexadecimal $04$ to $x$ and $y$, 
where $k.g=(x,y)$ picking $x,y \in [0,p-1]$ and writing them as $256$ in binary numbers. Usually a compressed form is used 
for public keys by hashing $K$ using SHA-256 then RIPEMD-160. This result is usually written in a special base 
called {\textit {Base58Check}}.

Since the public key and its compressed version are generated by applying one way cryptographic functions, one cannot recover 
the private key from them. Therefore they can be safely shared without compromising the private key. This is what 
is done when a payment is requested to an address.

The space of private keys is huge, thus if a proper random choice is done, there is no 
need to check that anyone else holds the same private key. This allows a decentralized generation of bitcoin 
addresses off-line and with no link to the identity of the owner.

This pseudanonymity plus the convenience of electronic payments has made bitcoin the currency of choice for payments made in the 
``deep web'', for example in Silk Road, the on-line market for drugs dismantled by the FBI.

This structure of bitcoin addresses raises many legal questions. The most basic one is how to prove ownership of bitcoins.
If someone wishes he can prove that he is in control of a certain address by signing a transaction (or any message) using his private key. 
On the other hand it is difficult to prove ownership of a private key with no collaboration.
%


\section{Why is bitcoin money? What is money?}\label{sec:money}

Bitcoin does not fit the standard definitions of money which can be found in most economy or finance books. Even some reputed 
economists have denied that bitcoin could be money. Before explaining why bitcoin is indeed good money we need to understand 
what the essence of money is. Some abstract reasoning is necessary.

The key to understand what money is relies in the fact that abstract objects exist \textit{only} on its properties. 
In other words, its existence does not depend on any
representation. Something which is familiar to us, mathematicians, is often difficult to 
grasp by non-mathematical minds which are used to rely on representations of the abstract objects. In a simple terms:

\textbf{Something with all the good properties of money is money!}

In layman's terms: If an animal looks like a duck, walks like a duck, swims like a duck, and squawks like a duck, then...it is a duck!
 
Now, what are these ``good properties'' of money?
Since prehistory, all civilizations noticed that barter was inefficient for trading, for the simple reason that most often the buyer has nothing of interest 
to offer to the seller in exchange for his goods, who may need other goods produced by a third individual. The solution to this game theory 
problem is to use an ``exchange token'' that 
has the consensus of the community in exchange of goods or services. Since ancient times all kinds of tokens have been used 
for that purpose: Shells, salt, seeds, metals, cigarettes, etc 

Obviously one accepts the token only if he is confident that in the future it 
will be accepted in exchange for goods or services preserving its value. Anything can be money, but \textbf{good money is confidence}. 
This monetary property makes good money valuable, a type of
value that transcends the physical representation it may have. Once this is understood, a natural list of properties that good money follows 
easily. No form of money has all these properties, and there is no universal notion of ``best money'' which depends on the context. 

\begin{enumerate}

 \item \textbf{Good money is not easy to falsify or produce.} Otherwise whoever has this capability could produce arbitrarily 
large quantities of money at minimal cost 
with the subsequent disruption of the monetary mass. In the case of gold, its chemical nature as an inert  fundamental element in Mendeleev Table makes it 
impossible to produce which was the dream of medieval alchimists 
(excluding residual decay of heavier elements from nuclear fission, which is not cost efficient). 
Shells could be used as money far away from the coast. 
Cash, coin and bills, are 
protected by security measures. Bitcoins are not forgeable because they are attached to unique bitcoin addresses which 
are linked criptographically in the unalterable blockchain to the original \textit{coinbase transaction} where they were created.

\item \textbf{Good money is easily authenticatable.} The success of gold is partly due to its properties (weight, color, 
texture) which makes it easily identifiable. In the case of bitcoins the blockchain data is used. 

\item \textbf{Good money is easily divisible.} Because we may need to buy cheap items and we should be able to fraction the payment received. Gold 
is easily divisible up to gram fractions. Euros are divisible into $100$ cents, and bitcoins into $100$ million
satoshis. 

\item \textbf{Good money is easily transportable.} Because we may use it far away from where we acquired it, 
for example if we migrate or if we trade internationally. 
Electronic money has the best transportaion properties. Gold is easy to transport only in small quantities, but silver is worse due to the higher mass/price ratio.

\item \textbf{Good money enables fast payment settlements.} It takes about ten minutes to validate a bitcoin transaction, and about 24 hours for an european banking SEPA.

\item \textbf{Good money is scarce.} There is no absolute measure of ``scarcity''. By ``scarce'' we mean that there must be some cost to its production or extraction. 
Otherwise its monetary value cannot exceed its producing cost. This explains why water cannot be good money. Gold is scarce because of the limited amount 
of gold that can be extracted from the Earth cortex, and it costs energy to extract it (mountains have literally been moved, as 
testifies the landscape of Las Medulas in Spain). 
Fiat money scarcity 
depends only on the will of the central banks (Quantitative Easing (QE) policies by the Fed makes the dollar abundant, but only to the financial sector). 
Bitcoin scarcity is encoded in the protocol: It costs energy to produce by the PoW and only $21$ million bitcoins will ever exist. Paradoxically, scarcity is not 
opposite with global use if divisibility is good. No more than 21 million people can have more than one bitcoin, but there are more than $200 \ 000$ satoshis for each citizen of the World.

\begin{figure}
\centering
\includegraphics[width=.8\linewidth]{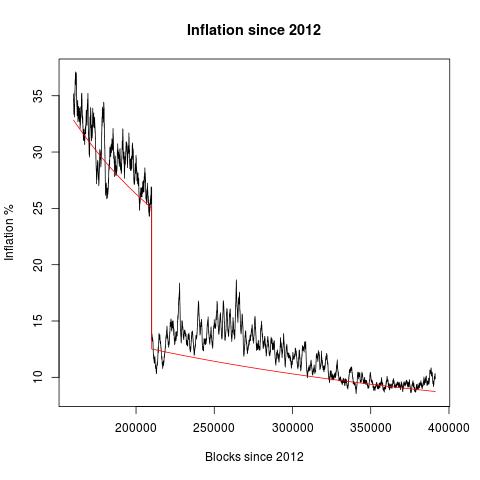}

\caption{Bitcoin price inflation since 2012.}\label{fig:inflation}
	
\end{figure}

\item \textbf{Good money is international.} Otherwise when we travel abroad we incur in exchange fees. Also the value of money attached to countries suffer 
from its economic situation. Gold is international and universally accepted. 
The US dollar is the most international national fiat currency. Bitcoin is transnational by design and not linked to any country or central bank.

\item \textbf{Good money preserves or increases its value over time.} \label{preserve}
We may need to save for years to make big purchases. Any sort of money suffers
fluctuations over time. Gold has passed the test of time: One gold ounce  
today and during the roman empire was worth an elegant suit. Historically fiat money is debased through inflation, in a way that 
``no one man in a million will detect'' (Keynes). Bitcoin has yet to pass the test of time. In 2011, 2012, 2013 and 2015 it was the best performing currency and in 2014 the worst one.

\item \textbf{Good money is not volatile.} Since we may not known when we would need to spend our savings, we want a stable value. All financial assets 
are volatile, but the free market exchange rate of bitcoin is highly volatile which is not desirable. This volatility is related to its increase 
of value which works by a concatenation of bubbles. Because of fundamental reasons volatility is decreasing over time. We will discuss this later.

\item \textbf{Good money is fungible.} This means that every dollar is like any other one, any gold atom is like any other one. 
Fungibility is not always perfect. For instance, money has a memory which will make it less fungible 
(for example fiat money coming from illicit activities needs to be laundered in order to be fungible). Bitcoin is not perfectly fungible 
since the blockchain keeps a trace of its past. Forinstance one can follow stolen bitcoins which are sold at a discount. 
Gold coins are not always fungible, since its preservation state determines part of its price even for bullion coins. 

\item \textbf{Good money does not decay over time.} Gold being chemically more neutral is better than silver. 
Treasures from wreckages give a good example: Gold coins can be in gem state, but silver coins are always badly damaged.

\item \textbf{Good money has a large base of users.} You only need one person to accept your money, but the larger is the community, the more efficient is the free 
market. 

\item \textbf{Good money is liquid.} It should be well accepted and exchangeable easily for other forms of money. 

\item \textbf{Good money is easy to store securely.} Since we may want to save it to spend it in the future, we must store it and protect it from thieves. Bitcoin
has some of the best storega properties.

\item \textbf{Good money is anonymous.}  Historically there is a right to privacy in comercial transaction,
and for security reasons anonymity is suitable for large transactions. Bitcoin is only pseudoanonymous. 

\item \textbf{Good money is decentralized.} Otherwise its value depends on a third party. 
If faith on this third party weakens, then confidence disappears 
and the monetary value is lost. This can happen suddenly. It represents an inherent unstability of fiat currency and derivatives markets. 
Centralized money can be blocked or confiscated by the central authority (for example in a bank freeze). 
Gold is decentralized, as bitcoin by design. 

\item \textbf{Good money is useless! } This is a major property that has been traditionally overlooked.
By ``useless'' we mean of non-monetary use: It better does not have any other 
significant use. In particular, no industrial use. 
Why must it be useless? Because otherwise part of its value, depends on the economic activity 
in the sector where it is employed. But according to point \ref{preserve}  its value must be preserved, and in particular be resilient to economic 
downturns. In those  cases good money should act as a refuge for wealth. When it has some other uses, it has also the positive effect that 
its value cannot collapse to absolute $0$, something which can happen to bitcoin, and happens too often with fiat money. 
When this money has no other use, its value is purely monetary. This property is the reason why silver, platinum and palladium 
have worse monetary value than gold. Platinum and palladium are used in the automobile industry in catalytic converters. 
Silver has been used in photography since the XIX-th century, and today is an essential element in 
the manufacturing of photovoltaic panels.

\begin{corollary*}
 \textit{The more abstract money is the better is. } 
\end{corollary*}

\begin{figure}
\centering
\includegraphics[width=.8\linewidth]{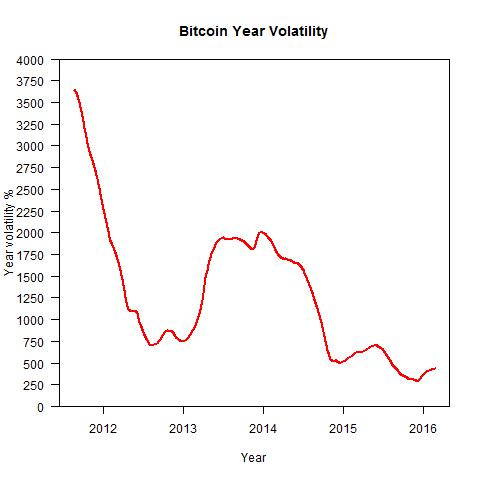}

\caption{Bitcoin volatility.}\label{fig:volatility}
	
\end{figure}

\item \textbf{Good money is antifragile.} According to Taleb's definiion \cite{Taleb2012}, something is antifragile if it gains from 
disorder and unexpected changes.
Good money is resilient to economic downturns, and also benefits from them as one of 
the best forms to preserve value. Bitcoin benefits from the inherent unstability of other centralized forms of money.

\item \textbf{Good money can be used over insecure channels.} Raw bitcoin transactions are designed to travel through insecure 
communication channels which is not the case of other digital payment methods that need an additional layer of encryption.

\item And last but not least, a bonus property new to bitcoin: \textbf{Good money is programmable!} 
This is a new property of bitcoin transactions: They come with 
an ``script field'' with instructions on how and when the transaction is to be performed. The 
most common script is to request that the funds can only be spent with the signature with the secret key of the destination address. 
But one can also specify things such as to request the transaction to be effective only at a later date, for example $52.560$ blocks later which 
is about one year later. Also the scripting allows $n-m$ multisignature addresses where $n$ signatures among $m\geq n$ are
necessary to spend the funds. 

\end{enumerate}

This last property is unique to bitcoin that is the first form of "`programmable money"'. Bitcoin is reinventing money. 
We conclude that bitcoin has many of the good properties of money, and therefore it is money! 

It is not ``perfect money''. Good properties of money depend on its use and context. John Nash gave a thought to what ``ideal money'' 
would be \cite{Nash2002}.
  
Nowadays its main weak points are the small users base and the high volatility. 
The reasons for high volatility are deep and fundamental: The distribution of bitcoins, concentrated in 
miners and early adopters, is heavily non-paretian. As observed by Pareto at the end of the XIX-th century, all wealth 
distributions have a power law decay, these distributions are universal and stable (\cite{PerezMarco2014} for a simple model 
explaining Pareto's distribution and stability, motivated by the study of bitcoin volatility). 
Therefore bitcoin wealth distribution must converge to a Pareto 
distribution and this induces an important volume in the market. High volume is correlated with high volatility.
From this we infer that volatility must be high, but also that it must decrease overtime which is what is being observed.

\section{New economic theory.}

Bitcoin is also a fascinating academic experiment: For the first time in history we can study the emergence of a new
decentralized monetary system. This will be a new chapter of its own in future textbooks in Monetary Theory.

Moreover it is destroying some carved in stone economic beliefs.  
Some economists have serious doubts about the viability of a deflationary currency. In simple terms their argument runs as follows:
Why anyone would be willing
to purchase anything if sometime later he can buy more with the same money? Their conclusions are that the consumption 
will vanish and the economy will perish. But historiacll gold has been mildly inflationary and viable form of money.
The only way to drive out gold from the monetary system has been by brute force by hoardind it massively by the central banks.

Actually, this scenario does not occur with bitcoin. Things do not work that way. With a deflationary currency 
there is no incentive to make superfluous expenses. People will purchase what they need and save the rest for 
the future. In practice what happens, as 
in the bimetallism dynamics, is that people will spend the bad money and hoard the good one, and also will spend bitcoins when 
its exchange rate is favorable.   

What does not work (in its traditional form) with a deflationary currency is the credit system at large. 
First, there is no incentive to place 
savings in a bank. One of the main reason why people do so is 
to mitigate the debasement of the currency with the interest paid by the bank. 
Loans are difficult to repay since on top of the deflation one has to pay interest. Only credits to very productive businesses
make sense. Mortgages are not. The banking foundations is at risk with a 
deflationary currency. This, together with the fact that bitcoins allow anyone to be its own bank, points to a profound 
disruption of the banking system, similar to what happen with the postal services with the general adoption of email.

Some economists believe that a currency without government sponsorship is hardly viable. 
This is the opinion of Yannis Varoufakis \cite{Varoufakis2013} a lucid 
contemporary economist, game theorist, and one of the few that understands the technical aspects of Bitcoin. 
Varoufakis had a brilliant plan to issue a Greek national cryptocurrency linked to the growth of the Greek GDP.

 \section{Regulatoy and legal status of bitcoin.}

Regulators are struggling to give bitcoin a proper legal status. One of the main questions is to decide if it should be 
considered a currency or a commodity. In the second 
case it would be subject to VAT, in the first case it will be exempted as other currencies (including bullion gold). The regulation is ot uniform.
Only recently the UE autorities have decided to consider bitcoin a currency. Probably it should not be consider neither, since it doesn't fit in the existing legal framework. In the USA, FinCen which supervises financial companies and enforces AML (Anti Money Laundering) regulations is requiring bitcoin exchanges to be regulated like financial service companies. A too strict regulation forces bitcoin businesses to migrate 
to more friendly jurisdictions.
Fiscal regulations problems are  also byzantine. What should the taxation of miners be? 
How can the fiscal rules be enforced with a pseudoanonymous currency? The ignorance by regulators of the 
technical aspects makes the task even more difficult. There have been some well thought reports by the ECB on virtual currencies \cite{ECB2012}. 
It is interesting to note how even the  ECB recognises their uneasiness 
with current definitions of money in the economic literature which do not seem to apply to bitcoin. From the introduction of the 2015 report:

\textit{The ECB does not regard virtual currencies, such as Bitcoin, as full forms of money as defined in economic literature. 
Virtual currency is also not money or currency from a legal perspective.}

 They even try their hand at giving a definition of virtual currency (in p.25):

\textit{A virtual currency can therefore be defined as a digital representation of value, not issued by a central bank, credit institution 
or e-money institution, which, in some circumstances, can be used as an alternative to money. }

An odd definition which defines as ``virtual currency'' anything not fitting the usual definition of money, missing the essence of 
what money really is: \textbf{Confidence}.

\section{The Trust Machine.}

Being an unfalsifiable ledger, the blockchain has other unsuspected applications. It has been rightly labeled as ``The Trust Machine'', which is 
a very accurate description. 

The first non-monetary application is to provide a decentralized trustable clock...by just counting blocks! It is not a highly 
precision clock because block validation occurs only in about $10$ minutes. 
Another example of new application is to provide notary services without notary! We can validate
the existence of a legal document directly inside the blockchain, and we don't need a third party. One can scan the document, hash it, 
and insert the hash in the transaction scripting 
space which will be encrusted in the blockchain and will remain there forever bringing mathematical proof that the document existed 
at the time of the block validation. Similarly, the blockchain can be used as a decentralized and universal system to certify proof of ownership.
We can include in the blockchain non-erasable messages as the one Satoshi inserted in the genesis block: ``The Times 03/Jan/2009 Chancellor 
on brink of second bailout for banks''.

The blockchain technology can be used in order to build decentralized and anonymous market places. Different projects are being developed. 
One can build other blockchains for this 
type of applications, or one can build what have been called ``side-chains''. This is a new idea which grafts a new service to the bitcoin blockchain, 
taking advantage of the security already in place for the bitcoin net. Also new platforms biuld from scratch as Ethereum are being proposed whose 
goal is to offer all sorts of decentralized services.

Since bitcoin offers the possibility of cheap and secure international transactions, it has attracted the attention of the banking system as a way to 
reduce their operational costs. There has been recently much talk about ``private'' blockchain, without realizing that the blockchain technology is about 
an open decentralized system and is more than a shared database. Some of this speculations fail to realize that a protocol of blockchain type  
is non-trivially linked to a currency token which incentives security by the decentralized mining power.

\section{Bitcoin curiosities.}

\subsection{Who is Satoshi Nakamoto?}

Little is known about the true identity of Satoshi Nakamoto, apart from his participation in the Cryptography Mailing List, 
a meeting place for the Cypherpunk group,
and in the forum \textit{bitcointalk} that he created. There has been much speculation. 
Even the name of the late John Nash was put forward, because of his writings on 
``Ideal Money'' \cite{Nash2002}. This is 
unlikely in view of the last mathematical section of the founding article, but his work could have had an influence. 
Nakamoto is an individual or a group of people, with a background in cryptography, 
mathematics and coding, leading the project up to their last post 
on December 2010. The diversity of expertise needed to device and implement the protocol, and the earlier precursor 
work,  points to a collaborative project by members of the cypherpunk community. This can explain also that early mined coins by Nakamoto
(estimated to be over a million) haven't moved. 
But why the Nakamoto group choosed to stay anonymous? 
When a project transcends individuals, it makes sense to develop it anonymously. We have a close example in the  
Bourbaki group (who may inspired them). The right to privacy is the most fundamental part of the Cypherpunk 
philosophy as described by the Cypherpunk manifesto (Erik Hughes, 1993). As for mathematical theorems, the identity of 
the authors is irrelevant for the ongoing project. We can only be grateful to them. 

\subsection{The millionaire pizza.}

The first bitcoin transaction was made in block $170$, on January 12th 2009, between Nakamoto and the late Hal Finney, who was the first one 
to realize the potential of Nakamoto's creation. The first purchase using bitcoin was only made much
later on the 22nd May 2010 \cite{bitcointalk2010}. The bitcointalk forum is a meeting place for bitcoin
developers and users where general and technical discussions take place. L. Hanyecz, the first GPU miner, offered to trade $10.000$ bitcoins for two pizzas
(which was worth at the time about \$40) to be delivered at his place. Another participant accepted the deal and earned what is now 
worth several million dollars. This gives a good illustration of the workings of a deflationary currency.

\subsection{Is Bitcoin a Ponzi scheme?}

Detractors of Bitcoin argue that the currency has nothing backing it, hence it has no value. 
They fail to grasp that Bitcoin is backed and secured by the full computer power 
of the validating network, currently at more than $9.4$ million petaFLOPS which at present surpasses hundreds of times the power of the $500$ fastest computers 
in the World combined together. Then they claim that Bitcoin is a mere Ponzi scheme, confusing its exponential viral development with a Ponzi scam. 
Viral development is common in the world of new technologies. The adoption rate is a sigmoid curve, with an exponential rate in the first phase (in which we are now). 
A main difference between bitcoin and a Ponzi scheme is the open source code. Anyone can check that the Bitcoin network 
functions as it is supposed to by inspecting and analysing the bitcoin code. Thus we have have full transparent information.
This is quite different from the opaque operative of Ponzi schemes (or the central banking system by the way).

\subsection{On bitcoin bubbles.}

The price of bitcoin is fixed by the free market. It is a non-regulated international market. Anyone can exchange other currencies or other goods for bitcoins if he finds 
a counterpart willing to accept bitcoins. The actual exchange rate is determined by the largests exchanges which have the largest volume. The price evolution 
is not peaceful. The bitcoin exchange rate suffers from high volatility. So far we have had 6 major bubbles, and many others at 
smaller scales: 
\begin{itemize}
 \item  From \$ $0.003$ to a peak of \$ $0.0875$, from May 1st to July 18th 2010, \textbf{2\ 816\%} increase in less than 3 months.  
 \item  From \$ $0.06$ to a peak of \$ $0.50$, from October 6th to November 6th 2010, \textbf{733\%} increase in one month.
 \item  From \$ $0.40$ to a peak of \$ $1.10$, from January 26th to February 9th 2011, \textbf{175\%} increase in two weeks.
 \item  From \$ $0.58$ to a peak of \$ $32.00$, from April 4th to June 8th 2011, \textbf{5\ 417\%} increase in two months.
 \item  From \$ $9.75$ to a peak of \$ $260$, from October 26th 2012 to April 10th 2013, \textbf{2\ 566\%} increase in about 5 months.
 \item  From \$ $110$ to a peak of \$ $1\ 245$, from October 2nd 2013 to December 5th 2013, \textbf{1\ 032\%} increase in 2 months.
\end{itemize}
These are partly speculative bubbles. Usually when these bubbles burst, the price does not drop significantly lower than the 
previous peak, which is an indicator that a fundamental upper trend is at work. Someone holding \$1 in bitcoin in April 2010 would have turned it today into 
\$$150\ 000$. The current anual monetary inflation is about 8\% (the real one, due to the increase of 
computing power is currently around 10\%), but the users growth is now higher. In 2016 after the next halving, inflation will drop at around 4\%.

\subsection{What is the target capitalization of bitcoin?}

Currently there are about $15$ million bitcoins (the 15th million bitcoin was mined on Christmas 2015). at the current market price of $400$ Euros  gives 
a capitalization of the bitcoin of about $6$ billion Euros. To put this number into perspective, a company enters in the list of the $50$ largest 
publicly traded companies with 
a market capitalization over $93$ billion Euros. Some small countries have a GDP smaller than bitcoins capitalization. 
The current monetary mass M1 of the Euro (currency in circulation + overnight deposits) is 
of $6 \ 500$ billion Euros (sea in order to operate with full functionalitysonally adjusted on October 2015 from ECB data).

One can speculate about the target capitalization of bitcoin. This depends on its growth and use. If only with a small fraction of the total 
underground economy goes into bitcoin its price will be quite large, \$ $10.000$ is not unreasonable, 
but some authors speculate that it may reach in a few years \$ $100 \ 000$ and even over $1$ million dollar per bitcoin if bitcoin 
becomes the reference international currency.

\subsection{What can I buy with bitcoins?}

There are all sorts of companies and small business accepting bitcoins. It is likely that somewhere near your place you may find some cafe or 
restaurant that accepts bitcoin. There is a useful site, coinmap.org, that plots in a map the places which accept bitcoin. One of the largest companies 
which sells goods over the Internet, Overstock, has been accepting bitcoins. 
Bitcoin is particularly well adapted to payments over the Internet. One does not 
need to worry about someone stealing your credit card data. When you make a bitcoin payment you don't give away any 
personal information nor any information that could be useful to a hacker. There are also companies, like Xapo, offering a bitcoin 
wallet linked to a regular credit card that can be refilled using bitcoins. 

\subsection{Can bitcoin be a solution for the unbanked?}

The major part of the population in the World has no access to banking services. In Kenya a particular form 
of electronic money developed in 2007. It is called \textit{mPesa} and the currency are cell phone minutes which can be exchanged and serve to pay for goods 
or services. About a third of Kenya population uses mPesa. Recently there have been efforts to integrate bitcoin with mPesa which will bring bitcoin 
to the users of mPesa which can be used with their regular cell phone. 

\subsection{Getting started.}

A final few lines on how to get started. The first question anyone asks is how to buy or earn some bitcoins. But first one needs to set up 
a wallet in order to manage bitcoins. The fastest way is to sign up for a free online wallet at sites 
like blockchain.info that also has an app for your android smartphone (you only need a working email). 
A better way is to download the official client at bitcoin.org, or a free wallet program like Armory or Electrum. The official client 
downloads the full blockchain. It can take several days (now the blockchain 
is more than 55 Gb heavy). Electrum avoids this by using ``trusted sites'' which is faster but not decentralized. Finally, for those wanting the highest level 
of security the best solution is to purchase a Trezor or Ledger device that is like a pendrive holding securely the private keys are stored that never go on-line. 
Then one can purchase or sell bitcoins by registering at an exchange as bitstamp.net or bitfinex.com (the two largests in the west), or do a personal transaction 
through the site localbitcoins.com.

\section{Conclusions.}

Monetary questions have historically attracted some of the most brilliant thinkers: Oresme, Copernicus, Newton, Keynes, Nash, Nakamoto...This new form of money 
which the Bitcoin protocol brings to us forces a redefinition of basic notions. It represents an escalation in the ladder of abstraction of the concept of money. 
As for mathematical theories, abstraction comes hand-to-hand with more pure, more general, 
more universal and more effective properties. From the perspective of Economic Dynamics this is a unique event which deserves close attention from the academic community. 
From the social and political point of view  bitcoin brings back to the citizen the ownership of money. 
It is a ``technology coup'' to the centralized outdated financial system ruled by banks.
The wide implementation of ``The Mathematical Trust Machine'' will have a profound social, economic and 
political impact, and Mathematics is at the root of this coming revolution.

\section*{Aknowledgements.}

I am grateful to L. Blanco, R. K. Englund, A. Marti  and N. Szabo for their remarks and comments.


\bibliographystyle{plain}
\bibliography{bitcoin}

\begin{thebibliography}{10}

\bibitem{Nakamoto08}
S.~Nakamoto, ``Bitcoin: A peer-to-peer electronic cash system,'' {\em
  www.bitcoin.org/bitcoin.pdf}, released on November 1st 2008 on the USENET
  Cryptography Mailing List "Bitcoin P2P e-cash paper".

\bibitem{Nakamoto09}
S.~Nakamoto, ``Bitcoin v0.1,'' released on January 9th 2009 on the USENET
  Cryptography Mailing List "Bitcoin v0.1 released".

\bibitem{CDLI}
``Cuneiform digital library iniciative,'' {\em www.cdli.ucla.edu}.

\bibitem{ServalTraine2014}
J.-F. Serval and J.-P. Trani\'e, {\em The Monetary System: Analysis and New
  Approaches to Regulation}.
\newblock Wiley \& Sons, 2014.

\bibitem{Flandreau2004}
M.~Flandreau, {\em The glitter of gold}.
\newblock Oxford University Press, 2004.

\bibitem{Copernicus1526}
N.~Copernicus, {\em Monete cudende ratio}.
\newblock 1526.

\bibitem{Oresme1360}
N.~Oresme, {\em De origine, natura, jure et mutationibus monetarum}.
\newblock 1360.

\bibitem{Taleb2012}
N.~Taleb, {\em Antifragile: Things that Gain from Disorder}.
\newblock Random House, 2012.

\bibitem{TheEconomist2015}
{\em The Economist}, vol.~417, 8962, October 31st-November 6th, 2015.

\bibitem{LAMPORT:1982}
L.~Lamport, R.~Shostak, and M.~Pease, ``The byzantine generals problem,'' {\em
  ACM Trans. Prog. Lang. Syst.}, vol.~4, pp.~382--401, July 1982.

\bibitem{github}
``Bitcoin code,'' {\em www.github.com/bitcoin}.

\bibitem{Nash2002}
J.~F. Nash, ``Ideal money,'' {\em Southern Econ. J.}, vol.~69, 1, 2002.

\bibitem{PerezMarco2014}
R.~Perez-Marco, ``A simple dynamical model leading to pareto wealth
  distribution and stability,'' {\em arXiv:1409.4857}, 2014.

\bibitem{Varoufakis2013}
Y.~Varoufakis, ``Bitcoin and the dangerous fantasy of apolitical money,'' {\em
  www.yanisvaroufakis.eu}.

\bibitem{ECB2012}
ECB, ``Virtual currency schemes,'' {\em www.ecb.europa.eu}, 2012, 2015.

\bibitem{bitcointalk2010}
L.~Hanyecz, ``Pizza for bitcoins!,'' {\em www.bitcointalk.org}.

\end{thebibliography}

\begin{info}
Ricardo P\'erez-Marco [\url{ricardo.perez.marco@gmail.com}] is Directeur de Recherches at CNRS, Paris, France.
\end{info}

\end{document}